\documentclass[aps,twocolumn,nofootinbib]{revtex4}
\usepackage{bbm}
\usepackage{mathrsfs}
\usepackage{amsmath}
\usepackage{graphicx}
\usepackage{dcolumn}
\usepackage{bm}
\usepackage{amssymb}
\usepackage{latexsym}

\begin{document}

 \title{Black Holes in the Universe: Generalized
Lema\^{\i}tre-Tolman-Bondi Solutions}

\author{Changjun Gao} \email{gaocj@bao.ac.cn}
\author{Xuelei
Chen} \email{xuelei@cosmology.bao.ac.cn} \affiliation{The  National
Astronomical Observatories, Chinese Academy of  Sciences, Beijing,
100012, China}
\author{You-Gen Shen}
\email{ygshen@center.shao.ac.cn} \affiliation{Shanghai Astronomical
Observatory, Chinese Academy of Sciences, Shanghai 200030, China}
\affiliation{Joint Institute for Galaxy and Cosmology of SHAO and
  USTC, Shanghai 200030, China}
\author{Valerio Faraoni}
\email{vfaraoni@ubishops.ca}
\affiliation{Physics Department, Bishop's University, 2600 College
Street, Sherbrooke, Qu\'{e}bec, Canada J1M~1Z7}

\date{\today}

\begin{abstract}
We present new exact solutions {which presumably describe} black
holes in the background of a spatially flat, pressureless dark
matter (DM)-, or dark matter plus dark energy (DM+DE)-, or
quintom-dominated universe. These solutions generalize
Lema\^{\i}tre-Tolman-Bondi metrics. For a DM- or (DM+DE)-dominated
universe, the area of the black hole apparent horizon (AH)
decreases with the expansion of the universe while that of the
cosmic AH increases. However, for a quintom-dominated universe,
the black hole AH first shrinks and then expands, while the cosmic
AH first expands and then shrinks. A (DM+DE)-dominated universe
containing a black hole will evolve to the Schwarzschild-de Sitter
solution with both AHs approaching constant size. In a
quintom-dominated universe, the black hole and cosmic AHs will
coincide at a certain time, after which the singularity becomes
naked, violating Cosmic Censorship.
\end{abstract}

\maketitle

\section{Introduction}

What is the relation between cosmic expansion and local
physics? Do local gravitational system, {\em i.e.}, stars,
galaxies, galaxy clusters, or even black holes, expand
with the cosmic expansion? This issue has received much
attention over the years
\cite{bon:1999,
nol:1998,pac:1963,noe:1971,sat:1983,sus:1985,
eat:1975,fer:1996, coo:1998,bol:2001,nol:1999,nol:1999-31,
bus:2003,gao:2004,sul:2005, miz:2005,nes:2004,pri:2005,
bal:2007,mas:2007,gao:2008}, however,
these discussions often produced contradictory results leaving
the issue rather ambiguous
(see \cite{kra:1997,sen:1999, CarreraGiulini} for
reviews). It is generally believed that these
contradictions  are due to  the use of different or
unphysical coordinates, or of unphysical solutions
\cite{coo:1998}.

In order to address this issue, McVittie introduced his renowned
solution \cite{mc:1933} in 1933; this {was intended} as describing
a point mass embedded in a Friedmann-Robertson-Walker (FRW)
universe, but modern studies revealed that the McVittie metric
{cannot} describe a point mass
\cite{sus:1985,nol:1998,nol:1999,mlc:2006}, but is rather
interpreted as describing a black hole in a cosmological
background  \cite{Kal:2010, lake:2011}. Einstein and Strauss
\cite{ein:1945} introduced the Swiss-cheese model which is,
however, unable to describe the Solar System and suffers from
other limitations \cite{bon:1999,kra:1997}. Next, the Vaidya
\cite{vai:1977}, Thakurta, Sultana-Dyer \cite{tha:1981, sul:2005},
and Faraoni-Jacques \cite{fa:2007} solutions were found, and these
{solutions also} describe  black holes embedded  in FRW universes,
each with some restrictions. Either the cosmic fluid is restricted
to be a cosmological constant (yielding the Schwarzschild-de
Sitter solution, a very special case, indeed), or it must be a
mixture of two perfect fluids, one of which is a null dust (for
the Sultana-Dyer solution), or it is an imperfect fluid \emph{with
heat flow} instead of a simple perfect fluid (for the
Faraoni-Jacques solution) \cite{gao:2008,mlc:2006}.  It is of
interest to look for simpler exact solutions with a single perfect
fluid as a source to address the issue of cosmic expansion versus
local physics. The purpose of the present paper is to derive such
new solutions representing  a black hole in a FRW universe with a
cosmic fluid consisting of dark matter, or of dark matter plus
dark energy (possibly a cosmological constant), or of quintom
matter \cite{feng:2004}. These types of solution are also useful,
if nothing else as toy  models, in the study of dynamical horizons
and their thermodynamics \cite{dynhorizons,
ward:2000,bak:99,cai:05,recent:2008} and of  primordial black
holes \cite{zel:1967} as probes of the early universe
\cite{har:2005,sai:2007}.

Regarding the relation between cosmic expansion and local physics,
Noerdlinger and Petrosian conclude their work \cite{noe:1971}  by
stating: ``\emph{Consider the possible expansion of clusters or
superclusters of galaxies, of mean rest-mass density $\rho_c$,
immersed in a universe containing a gas of particles having energy
density $\rho$. It is shown that when $\rho>\rho_c$, the clusters
or superclusters expand with the universe, but  if $\rho_c \gg
\rho$, the expansion is reduced in the ratio $\rho/\rho_c$}".
Unfortunately, this conclusion is drawn from the McVittie
solution, which can not describe {realistically} a point mass, a
star, or a cluster of galaxies in a FRW background. A similar
conclusion was reached  by Price \cite{pri:2005} who discovered
the ``all or nothing"  behavior ({\em i.e.}, weakly coupled
systems are comoving while strongly coupled ones resist the cosmic
expansion). However, Price's result applies only to a de Sitter
background, not to a general FRW universe. As is well-known, de
Sitter space is very special: the de Sitter metric can be put in
static form {in the spacetime region between the  black hole and
the  cosmological horizons}, which explains why strongly bound
systems in this background do not expand \cite{fa:2007}.

In this paper, we present new solutions {which we propose to
interpret} as describing cosmological black holes embedded in FRW
backgrounds, with  the advantage over previous solutions
\cite{mc:1933, vai:1977,  tha:1981, sul:2005, fa:2007} that {they
are} sourced by a \emph{single  perfect fluid}, without heat flux.
These solutions lead to surprising results: the black hole AH
decreases with the expansion of the universe when the latter is
DM- or (DM+DE)-dominated. Since the Misner-Sharp mass is a half of
the radius of the AH, also the black hole mass decreases with the
cosmic expansion. As the most  strongly bound gravitational
system, it is intuitive that the size of a black hole should
increase with the expansion of the universe due to the swallowing
of surrounding cosmic matter. However, this picture is incorrect
in general as the cosmic expansion ``wins'' over the local
gravitational attraction of the black hole. The universe is always
expanding after the Big Bang and the energy density of the cosmic
fluid decreases with cosmic time. The seemingly bizarre behavior
of the  Misner-Sharp mass derives from the fact that it is really
a mass sum ({\em  i.e.}, the mass of the background fluid is
included), {and coincides with the Hawking-Hayward quasi-local
mass} \cite{gao:2008}. We note that this is a purely classical
effect which makes the black hole mass decrease and has nothing to
do with Hawking radiation \cite{haw:1971}.

In a quintom-dominated universe, during the matter era the black
hole mass decreases, while during the phantom-dominated epoch it
increases. The physical reason is that the cosmic density first
decreases and then increases. Since the {Misner-Sharp (MS)} mass
is the sum of the black hole mass and of the background mass, the
black hole mass first decreases and then increases. One
interesting result is that the black hole singularity will become
naked before the Big Rip occurs, which violates the Cosmic
Censorship Conjecture \cite{pen:1969}. So if the latter is
correct, phantom matter may not exist in {n}ature.

In this paper, we use units in which  the speed of light $c$ and
Newton's constant $ G$ assume the value unity, and the signature
$+2$ for the metric. The paper is organized as follows: first  we
show that several known solutions can not describe a black hole
embedded in a matter-dominated universe. Second, we  present new
solutions describing a black hole in a matter-dominated universe, a
dark matter plus cosmological constant-dominated universe, and a
quintom-dominated universe. The last section contains the
conclusions.

\section{The McVittie solution}

A natural starting point to investigate the relation between cosmic
expansion and local physics is the 1933 McVittie solution
\cite{mc:1933},  which can be
written in isotropic coordinates as
\begin{eqnarray}
 \label{eq:DS-dt}
  ds^2&=&-\frac{\left[1-\frac{M_0}{2a\left(t\right)r}
\right]^2}{\left[1+\frac{M_0}{2a\left(t\right)r}\right]^2}dt^2
  +{a\left(t\right)^2}\left[1+\frac{M_0}{2a\left(t\right)r}
  \right]^4   \nonumber\\
&&\nonumber \\
  &&\cdot\left(dr^2+r^2d\Omega^2 \right)\;,
 \end{eqnarray}
where the constant $M_0$ reduces to the physical black hole mass
when $a(t)=$const. (this can be seen by rescaling the radial
coordinate $r\rightarrow \tilde{r}\equiv a{r}$), and the scale factor
$a(t)$ is an arbitrary function
of the cosmic time $t$. When $M_0=0$, the solution reduces to the
spatially flat FRW metric, and at
a first glance, it  can be understood as representing a
Schwarzschild black hole embedded in a spatially flat
FRW universe \cite{noe:1971}. However,
 it is now known that, with the exception of the Schwarzschild-de Sitter
solution, {matters are more complicated}
 because the McVittie metric is
singular on the 2-sphere $r=2M_0a$ (which reduces to the
Schwarzschild horizon if $a=$const.) \cite{sus:1985} and this
singularity is spacelike \cite{nol:1999}. It was  claimed in the past that
the McVittie metric  describes a point mass located at $r =0$
and embedded in a FRW universe. However, this point mass is, in general,
surrounded by the singularity at $r = M_0/(2a)$.  This singularity
was studied
in Ref.~\cite{sus:1985}.
Nolan \cite{nol:1998} showed that it is a weak singularity in the
sense that an object falling onto the $r = M_0/(2a)$
surface is not shrunk to zero volume, and therefore the energy density of
the surrounding fluid is finite. However, the pressure diverges at
$r = M_0/(2a)$ together with the Ricci scalar $R$
\cite{sus:1985,mlc:2006}. While there is little doubt that the
McVittie metric represents some kind of strongly gravitating central
object, its physical interpretation is not completely clear and is
still a subject of debate \cite{sus:1985, nol:1999,
Kal:2010, lake:2011}.

In view of this situation we show that, {with the exception of the
Schwarzschild-de Sitter solution}, \emph{the McVittie solution is
{unable to describe a black hole or  a point mass in a FRW
universe}  when the equation of state {of the cosmic perfect
fluid} is $p=w(t)\rho$}. The proof of this statement proceeds as
follows. The McVittie solution was obtained by forbidding
 explicitly the accretion of
cosmic fluid onto the central object. This requirement corresponds to
$G_0^1=0$, which in turn implies that the stress-energy tensor component
$T_0^1=0$ and there is no radial flow of energy. Let us assume that the
energy-momentum tensor that sources the McVittie metric is that of a
perfect fluid,
$T_{\mu\nu}=\left(\rho+p\right)U_{\mu}U_{\nu}+pg_{\mu\nu}$, where
$U_{\mu}$ is the 4-velocity of the comoving observer, and  $\rho$ and
$p$ are the density and pressure of the fluid, respectively.

By substituting the metric and the energy-momentum tensor into the
Einstein
equations $G_{\mu\nu}=8\pi T_{\mu\nu}$, the resulting Einstein
equations can be arranged as
 \begin{eqnarray}
 \label{eq:h-rho}
3H^2&=&8\pi\rho\;,
\end{eqnarray}
\begin{eqnarray}
\label{eq:ddot-a} \frac{2\ddot{a}}{a}+H^2&=&-8\pi
p-\frac{M_0}{ar}\left(\dot{H}-\frac{3}{2} \, H^2-4\pi
p\right)\;,
\end{eqnarray}
where
$H=\dot{a}/a$ is the Hubble parameter  and an overdot denotes
differentiation with respect to
the cosmic time $t$. We recognize eqs.~(\ref{eq:h-rho}) and
(\ref{eq:ddot-a}) as the Friedmann and the acceleration
equations, respectively.

We have $p=0$ and $p=\rho/3 $ for the matter-dominated
 and the radiation-dominated eras, respectively, while $p=w(t)\rho$
with $w<-1/3$ for  a
dark energy-dominated universe.  For all these cases, the equation
of state assumes the form $p=w(t)\rho$.
Eq.~(\ref{eq:h-rho}) tells us that $\rho$ depends only on $t$ and,
since $p=w(t)\rho$, the same is true  for
$p$. In eq.~(\ref{eq:ddot-a}), the expressions
$\frac{2\ddot{a}}{a}+H^2$ and $-8\pi p$ depend  only
on $t$, while $-\frac{M_0}{ar}\left(\dot{H}-\frac{3}{2}
\, H^2-4\pi
p\right)$ depends on both $t$ and $r$, therefore
eq.~(\ref{eq:ddot-a}) is
satisfied if and only if
\begin{eqnarray}
 &&3H^2=8\pi\rho\;,\ \ \
 \frac{2\ddot{a}}{a}+H^2+8\pi p=0\;,\nonumber\\&&
\dot{H}-\frac{3}{2} \, H^2-4\pi p=0\;.
\end{eqnarray}
These admit the unique solution
\begin{eqnarray}
H=\mbox{const.}\;, \ \ \ \ p=-\rho=\mbox{const.}\;, \ \ \ \
a(t)=\mbox{e}^{Ht}\;,
\end{eqnarray}
which is the Schwarzschild-de Sitter spacetime. Therefore,
\emph{the McVittie solution can not describe a black hole embedded
in a matter-dominated, radiation-dominated,  or dark
energy-dominated universe with {perfect fluid} equation of state
$p=w(t)\rho$.} In one word, given the equation of state
$p=w(t)\rho$, the McVittie solution is nothing but the
Schwarzschild-de Sitter solution.

Actually, Kaloper {\em et al.} \cite{Kal:2010} have argued that in
the presence of a cosmological constant, McVittie's solution does
in fact represent a black hole in an expanding universe. Here we
{provide independent support for this statement}.

{Recently,} assuming a specific function $H(t)$ in the McVittie
solution, Lake and Abdelqader \cite{lake:2011} performed a
detailed study {of} the particular McVittie solution and
corroborated the conclusion of Kaloper {\em et al.} {Moreover,}
using a tetrad-based method for solving Einstein's field
equations, Nandra {\em et al.} \cite{nandra:2011} obtained
solutions {describing objects embedded in FRW universes which
provide a new perspective on the McVittie metric}. {The study of
the global structure of the McVittie solution in \cite{lake:2011}
revealed unexpected features which make this solution much more
complicated than it would seem at a first sight. Something similar
probably happens for the new solutions presented here, although we
will leave the detailed investigation of radial null geodesics and
of the global structure to future work.}

\section{The Faraoni-Jacques solution}

Because the McVittie solution can not describe a black hole in a FRW
universe, we turn our attention to the Faraoni-Jacques solution
\cite{fa:2007}, which is a generalization of the McVittie solution
\begin{eqnarray}
  ds^2&=&-\frac{\left[1-\frac{M\left(t\right)}{2a\left(t\right)r}\right]^2}{\left[1+\frac{M\left(t\right)}{2a\left(t\right)r}\right]^2}dt^2
  +{a\left(t\right)^2}\left[1+\frac{M\left(t\right)}{2a\left(t\right)r}
  \right]^4
   \nonumber\\
&&\nonumber \\
  &&\cdot\left(dr^2+r^2d\Omega^2 \right)\;,
 \end{eqnarray}
in which the function $M(t)$ replaces the  constant $M_0$ of the
McVittie metric. This corresponds to lifting the McVittie
non-accretion restriction $G^1_0=0$. This solution {seems to
describe} a cosmological black hole embedded in a FRW background
and presents advantages over the previous Thakurta and
Sultana-Dyer solutions \cite{tha:1981, sul:2005},
\begin{eqnarray}
  ds^2&=&-\frac{\left(1-\frac{M_0}{2r}\right)^2}{\left(1+\frac{M_0}{2r}\right)^2}dt^2
  +{a\left(t\right)^2}\left(1+\frac{M_0}{2r}
  \right)^4
   \nonumber\\
&&\nonumber\\
  &&\cdot\left(dr^2+r^2d\Omega^2 \right)\;,
 \end{eqnarray}
in the sense that both the energy density and the pressure are
finite near the black hole horizon, and the energy density is
positive-definite \cite{fa:2007}. Now let us examine whether it can
describe a black hole embedded in FRW universe sourced by a perfect fluid.

We assume that the cosmological matter is described by the
single perfect fluid energy-momentum tensor
\begin{eqnarray}
T_{\mu\nu} = \left( \rho+p \right)U_{\mu}U_{\nu} +p g_{\mu\nu}\;,
\end{eqnarray}
where $U^{\mu}$ is the 4-velocity of the fluid. The only
non-vanishing
components of the Einstein tensor are
\begin{eqnarray}
 \label{eq:G-T}
  &&G_0^0\neq 0\;, \ \ \  G_0^1\neq 0\;, \nonumber\\
&& G_1^1=G_2^2=G_3^3\neq 0\;.
 \end{eqnarray}
If a radial energy flow is allowed, the fluid four-velocity is
\begin{eqnarray}
\label{eq:4vel0} U^{\mu}= \left(U^{0}, U^{1}, 0, 0\right)\;,
\end{eqnarray}
and the normalization $U_{\mu}U^{\mu}=-1$ gives
\begin{eqnarray}
\label{eq:4vel} U^{\mu}=
\left[\sqrt{{{-g^{00}}}-{{a^2v^2}}{g^{00}}{\left(1+
\frac{M\left(t\right)}{2a\left(t\right)r}\right)^4}}\;,
v, 0, 0\right],
\end{eqnarray}
where $v$ is the proper 3-velocity of the fluid. From
$G_{11}=G_{22}$ we conclude that
\begin{eqnarray}
\left(\rho+p\right)a^2v^2\left(1
+\frac{M\left(t\right)}{2a\left(t\right)r}\right)^4=0\;.
 \end{eqnarray}
We have two types of solutions characterized by
\begin{eqnarray}
v=0\;,\ \ \ \ M=M_0
 \end{eqnarray}
(which is the McVittie solution), or
\begin{eqnarray}
p=-\rho\;,
 \end{eqnarray}
 which gives the Schwarzschild-de Sitter solution.
Therefore,  \emph{the Faraoni-Jacques solution (and also
the Thakurta and Sultana-Dyer solutions) can not describe a black hole
embedded in a matter-dominated universe. In general, they can not
describe a black hole embedded in a single perfect fluid-dominated
universe}. However, the Faraoni-Jacques solution can describe a black
hole embedded in FRW Universe provided that the source is given by an
imperfect fluid with a heat flux:
\begin{equation}
T_{\mu\nu}=\left( p+\rho \right) U_{\mu}U_{\nu} +p
g_{\mu\nu}+q_{\mu}U_{\nu} +q_{\nu}U_{\mu} \;,
\end{equation}
where $U^{\mu}=\left( \sqrt{-g^{00}}, 0,0,0 \right)$ and
$q^{\mu}=\left(0, q ,0,0 \right)$, where $q$ is the heat flux
density. This has been shown by us in \cite{gao:2008}.

\section{The Vaidya solution}

The Vaidya metric is obtained by  conformally transforming
the Minkowski one:
\begin{eqnarray}
 \label{eq:vaidya}
  ds^2&=&a^2\left( \eta \right) \left(
-d\eta^2+dr^2+r^2d\theta^2+r^2\sin^2\theta
d\phi^2\right)
  \nonumber\\&&+\frac{2M_0}{r}\left(d\eta +dr\right)^2\;,
 \end{eqnarray}
where $M_0$ is the black hole mass and $\eta$ is the conformal time. When
$M_0=0$ and
$a=\eta^{2}$, this metric reduces to that for a  matter-dominated
universe. On the other hand, when $a=$const., it reduces to
the Schwarzschild metric. At a first glance, one  is tricked again into
believing that this metric may describe a black hole in a matter-dominated
universe but this is incorrect, as shown below.

The energy-momentum tensor of dust or dark  matter is
\begin{eqnarray}
 \label{eq:dm}
 T_{\mu\nu}&=&\rho_d U_{\mu}U_{\nu}\;.
 \end{eqnarray}
Due to spherical symmetry, the four-velocity $U^{\mu}$ is
written as eq.~(\ref{eq:4vel0}).
We have two vanishing components of $T_{\mu\nu}$, {\em i.e.}, $T_2^2$
and $T_3^3$, hence  $G_2^2=G_3^3=0$. Given the
metric~(\ref{eq:vaidya}),  $G_2^2=G_3^3$ is calculated as
\cite{mlc:2006},
\begin{eqnarray}
 \label{eq:g22}
\left( \frac{2\ddot{a}}{a^3}-
\frac{\dot{a}^2}{a^4} \right)+ \left( \frac{2m\ddot{a}}{ra^5}
-\frac{4m\dot{a}^2}{ra^6} \right)&=&0\;.
 \end{eqnarray}
The terms in the first bracket depend only on time,
while those in the second bracket are $r$-dependent,  so
eq.~(\ref{eq:g22}) is satisfied  if and only if
\begin{eqnarray}
 \label{eq:g33}
 \frac{2\ddot{a}}{a^3}-\frac{\dot{a}^2}{a^4}=0\;,\ \ \ \
 \frac{\ddot{a}}{a^5}-\frac{2\dot{a}^2}{a^6}=0\;.
 \end{eqnarray}
Then we necessarily have the trivial solution $a=$const., which is
just the
Schwarzschild metric. This shows that \emph{the Vaidya solution can
not describe a black hole embedded in a matter-dominated
universe}, either.

Thus far, we have shown that several known exact solutions can not
describe a black  hole in a matter-dominated universe. In the next
section, we look for such a  solution.

\section{Black hole in a matter universe}

{\bf A} spherically symmetric inhomogeneous gravitational
field produced by dark matter with
energy-momentum~(\ref{eq:dm}) is described by
the \emph{Lema\^{\i}tre-Tolman-Bondi} (LTB) metric \cite{bo:1947}
which, in the notation of
\cite{ce:2000}, is given by
\begin{equation}
\label{eq:TB} ds^2=-dt^2+\frac{R'^2}{1+f} \,dr^2+R^2d\Omega^2\;.
\end{equation}
Here $f$ is an arbitrary function of the comoving coordinate $r$
satisfying $f>1$, $R(t, r)$ is the physical radius at time $t$ and
coordinate radius $r$, while a prime represents differentiation  with
respect to $r$.

With the energy momentum tensor~(\ref{eq:dm}) and the
metric~(\ref{eq:TB}), the Einstein equations read
\begin{eqnarray}
\label{eq:EE1}  \frac{3\left( \dot{R}^2-f\right) }{R^2}=\frac{8\pi F}{R^3}\;,
\end{eqnarray}
\begin{eqnarray}
\label{eq:EE2} \frac{F'}{R^2R'}=3\rho_d\;,
\end{eqnarray}
where an overdot and a prime represent partial differentiation with
respect to $t$ and $r$, respectively, and $F(r)$ is an arbitrary
function of $r$. If we substitute
\begin{eqnarray}
\label{eq:fF}f=-f_0r^2\;,\ \ \ F=\rho_0r^3\;,\ \ \ \
R=a\left(t\right)r\;,
\end{eqnarray}
into eqs.~(\ref{eq:EE1}) and ~(\ref{eq:EE2}), the latter
become
\begin{eqnarray}
\label{eq:EE11}
\frac{3\left( \dot{a}^2+f_0 \right)}{a^2}=\frac{8\pi\rho_0}{a^3}\;,\ \ \
\rho_d=\frac{\rho_0}{a^3}\;,
\end{eqnarray}
with $f_0$ and $ \rho_0$  constants. We recognize
eq.~({\ref{eq:EE11}}) as the Friedmann equation with
a dust source, hence $f_0$ represents the spatial curvature
and $\rho_0$ the density of the universe at time $a_0=1$.
The observed universe is spatially flat with good accuracy (probably
due to inflation) and  we can  neglect the curvature term in what
follows by setting $f=0$.

For an arbitrary function  $F(r)$ and $f=0$, the solution of
eq.~(\ref{eq:EE1}) was given by Tolman \cite{to:1934} and
Oppenheimer
and Snyder \cite{op:1939} in their pioneering investigations  of
the gravitational collapse of dust as
\begin{eqnarray}
\label{eq:TOS}
R=\left[h\left(r\right)+\frac{3}{2}
\sqrt{\frac{8\pi}{3}F} \, t\right]^{2/3}\;,
\end{eqnarray}
where  the integration ``constant" $h$ is an arbitrary function of
$r$. Substitution of eq.~(\ref{eq:TOS}) into~(\ref{eq:TB})
yields
\begin{equation}
\label{eq:TB1}
ds^2=-dt^2+\frac{1}{R}\left(dh+\sqrt{6\pi}  \, t d \left( \sqrt{F} \right)
\right)^2+R^2d\Omega^2\;.
\end{equation}
There is actually less freedom in eq.~(\ref{eq:TOS}) than
is apparent
from the two arbitrary  functions $F$ and $h$ since $F$ can be taken
as a new radial variable. It is convenient  to choose
\begin{equation}
\label{eq:h} h=r^{3/2}\;,
\end{equation}
and
\begin{eqnarray}
\label{eq:TOS1} R=\left[r^{3/2}+
\frac{3}{2}\sqrt{\frac{8\pi}{3}F} \,t\right]^{2/3}\;.
\end{eqnarray}
Eq.~(\ref{eq:TOS1}) is the general solution representing a
spherically
symmetric gravitational field with  dust as a source. As
shown above,  this can describe a dust-dominated universe. On the
other hand, if
the dust disappears, it describes a vacuum, spherically
symmetric, gravitational field and reduces to the
Schwarzschild metric due to the Jebsen-Birkhoff
theorem. By combining these considerations, one guesses that
eq.~(\ref{eq:TOS1}) may describe a Schwarzschild black hole
immersed
in a dust-dominated universe. In the following, this claim is examined.

\subsection{Minkowski space}

By solving the Einstein equations with $\rho_d=0$, one finds the vacuum
solution
\begin{eqnarray}
\label{eq:vacuum} F=F_0\;,
\end{eqnarray}
where $F_0$ is an integration constant. If $F_0=0$, we
obtain $R=r$ and we recover the Minkowski solution.

\subsection{The Schwarzschild solution}

If $F_0\neq 0$, we recover the Schwarzschild solution. It is
convenient to set $F=\frac{3F_0^2}{8\pi }$ with $F_0$ a positive
constant. Then, the Schwarzschild solution is given by
\begin{eqnarray}
R=\left[r^{3/2}+\frac{3}{2} \, F_0t\right]^{2/3}\;,
\end{eqnarray}
$F_0$ is determined by the mass $m$ of the black
hole. In order to show this,  let us rewrite the metric in Schwarzschild
coordinates, setting
\begin{eqnarray}
\label{eq:SCH} x(t,r)
=\left[r^{3/2}+\frac{3}{2} \, F_0t\right]^{2/3}\;.
\end{eqnarray}
Then, eq.~(\ref{eq:TB}) becomes
\begin{equation}
ds^2=-\left(1-\frac{F_0^2}{x}\right)dt^2+dx^2
-\frac{2F_0}{\sqrt{x}}\, dt \, dx+x^2d\Omega^2\;.
\end{equation}
Using the new time variable
\begin{equation}
\label{eq:Tt}
T=t-t_0+2F_0\sqrt{x}+F_0^2\ln\left(\frac{\sqrt{x}-F_0}{\sqrt{x}+F_0}\right)\;,
\end{equation}
one obtains
\begin{equation}
\label{eq:TB2}
ds^2=-\left(1-\frac{F_0^2}{x}\right)dT^2
+\left(1-\frac{F_0^2}{x}\right)^{-1}dx^2+x^2d\Omega^2\;,
\end{equation}
{\em i.e.}, the Schwarzschild solution. The physical
meaning of $F_0$ is derived from $F_0=\sqrt{2m}$ and the
Schwarzschild solution is given by
\begin{eqnarray}
\label{eq:SCH1} R=\left[r^{3/2}+\frac{3}{2}\sqrt{2m} \,
t\right]^{2/3}\;.
\end{eqnarray}

\subsection{Dust-dominated universe}

Eqs.~(\ref{eq:EE2}) and~(\ref{eq:TOS1}) show that
the homogeneous and
isotropic, spatially flat, dust-dominated FRW solution is recovered if
and only if $F=\rho_0 r^3$.  The dust
universe solution is then given by
\begin{eqnarray}
\label{eq:dust}
R=\left[r^{3/2}+\frac{3}{2}\sqrt{\frac{8\pi\rho_0 }{3}} \,
r^{3/2}t\right]^{2/3}\;.
\end{eqnarray}
The scale factor and the density are
\begin{eqnarray}
a=\left[\sqrt{6\pi\rho_0} \,t+1\right]^{2/3}
\end{eqnarray}
and
\begin{eqnarray}
\label{eq:dustdensity} \rho_d=\frac{\rho_0}{a^3}\;,
\end{eqnarray}
respectively, and it must be $t\geq t_0$, with
$t_0=-\frac{1}{\sqrt{6\pi\rho_0}}$, representing the Big Bang.

\subsection{Black hole in a dust-dominated universe}

By combining the considerations above, one expects the metric for a black
hole immersed in a dust universe to be given by
\begin{eqnarray}
\label{eq:bhdust}
R=\left[r^{3/2}+\frac{3}{2}\sqrt{2m} \, t
+\frac{3}{2}\sqrt{\frac{8\pi\rho_0
}{3}} \, r^{3/2}t\right]^{2/3}\;.
\end{eqnarray}
If $m=0$, this line element describes a dust universe, while if
$\rho_0=0$, it reduces to a Schwarzschild solution;
(\ref{eq:bhdust}) {should describe} a black hole immersed in dust
universe. It is clear, by comparison of eqs.~(\ref{eq:bhdust})
and~(\ref{eq:TOS}), that this is a solution of the full Einstein
equations. The substitution of eq.~(\ref{eq:bhdust}) into
eqs.~(\ref{eq:EE1}) and~(\ref{eq:EE2}) gives the {energy} density
\begin{eqnarray}
\label{eq:density}
\rho_d=\frac{\sqrt{\rho_0}\left(3\sqrt{m}
+2\sqrt{3\pi\rho_0}r^{3/2}\right)}{2
\sqrt{3\pi}\left(\sqrt{6\pi\rho_0} \, t+1\right)R^{3/2}}\;.
\end{eqnarray}
The physical radius $R$ cannot be negative, hence
 we require that $r\geq0$ and $t\geq
t_0\equiv-\sqrt{{6\pi\rho_0}}$. Then, the density is always positive.
 If $m=0$, we have $\rho_d=\rho_0/R^3$, {\em i.e.}, dust or dark
matter. Since $R=0$ represents the Big Bang,  the latter occurs at
the time
\begin{eqnarray}
\label{eq:BBB} t=-\left(\frac{3}{2}\sqrt{2m} \,
r^{-3/2}+\sqrt{{6\pi\rho_0}}\right)^{-1}\;,
\end{eqnarray}
and the energy density is positive everywhere on the
spacetime manifold.

\section{Evolution of the apparent horizons}

In the Schwarzschild-de Sitter spacetime, there exist a
black hole apparent horizon (AH) and a cosmic AH. The
metric~(\ref{eq:TB})
describes a spherically symmetric and dynamical black hole in a FRW
background more general than de Sitter. In this case, the event
horizon may not be well-defined, but the AH always exists. The AH is
a marginally trapped surface with vanishing expansion and has been
argued to be a causal horizon for a dynamical spacetime. The AH is
associated with the Hawking temperature, gravitational entropy and
other thermodynamical aspects
\cite{ward:2000,bak:99,cai:05,recent:2008}. The first law of
thermodynamics for the AH has been derived not only in general
relativity but also in several other theories of gravity including
the Lovelock, nonlinear, scalar-tensor, and braneworld theories
\cite{gong:07,ak:07,caic:07,caicao:07,akc:07,she:07,shey:07}. In
view of this point, in order to investigate the evolution of the
black hole mass, we calculate the energy contained inside the AH
which has been used in black hole thermodynamics in relation with
energy flows through the AH \cite{gong:07, DiCriscienzoetal07}. Let us
proceed to calculate the radius of the black hole AH.

For a spherically symmetric spacetime with line element $ds^2
=h_{\mu\nu}dx^{\mu}dx^{\nu}+ x^2d\Omega^2$, the generalized
Misner-Sharp mass is $M_{MS}=x\left( 1- h^{\mu\nu} \,x_{,\mu}
x_{,\nu}\right) /2$ \cite{ms:64}. At the AH, it is $ h^{\mu\nu} x_{,\mu}
x_{,\nu} = 0$ and the generalized
Misner-Sharp mass inside the AH is simply
\begin{equation}
 \label{eq:MR} M_{MS}={x_{AH}}/2 \;,
\end{equation}
where $x_{AH}$ is the radius of the AH. In order to
know the evolution of the black hole MS mass, we need to
know the AH radius.

\begin{figure}
\includegraphics[width=6.5cm]{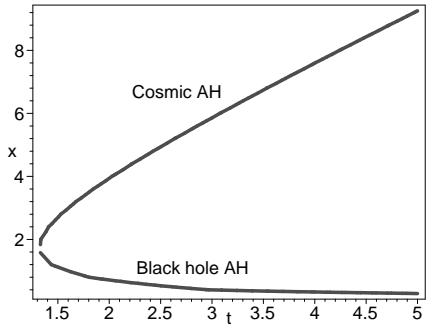}
\caption{The size of the black hole AH (lower {curve}) in a
dust-dominated universe decreases with the cosmic expansion  while
that of the cosmic AH (upper {curve}) increases. The  MS mass of
the black hole tends to zero in the future. There was a time at
which the two horizons coincided and before which they were absent
and the singularity was naked. The plots correspond to the
parameter values $ m=1, \rho_0=0.05$ in {Plank units}. In {the}
International  Units System, they are
$m=10^{24}~\textrm{M}_{\odot}$ and $ \rho_0=0.05\cdot
10^{-123}~\rho_P$ ($\rho_P$ is the Plank  density).}
\label{fig:x-t-dust}
\end{figure}

We rewrite eq.~(\ref{eq:TB}) in
Schwarzschild coordinates using the new spatial coordinate
\begin{eqnarray}
\label{eq:XXXX}
x=\left[r^{3/2}+\frac{3}{2}
\sqrt{2m}t+\frac{3}{2}\sqrt{\frac{8\pi\rho_0
}{3}}r^{3/2}t\right]^{2/3}
\end{eqnarray}
instead of  $r$, in terms  of which eq.~(\ref{eq:TB}) is
rewritten as
\begin{eqnarray}
\label{eq:sch}
&&ds^2=-\left[1-\frac{1}{x}{\left(\sqrt{2m}+ \sqrt{\frac{8\rho_
0\pi}{3}}r^{3/2}\right)^2}
\right]dt^2+dx^2\nonumber\\
&&\nonumber\\
&&
-\frac{2}{\sqrt{x}}\left(\sqrt{2m}+\sqrt{\frac{8\rho_0\pi}{3}}
r^{3/2}\right)dtdx+x^2d\Omega^2\;.
\end{eqnarray}
The coordinate system $ \left( t,x ,\theta,\phi \right)$ is not
orthogonal but the cross-term $dtdx $ can be eliminated by introducing
the new time $T$ defined by
\begin{eqnarray}
\label{eq:TT}
&&d{T} = \frac{1}{J\left(t, x\right)}
\left\{dt+\frac{1}{\sqrt{x}}\left(\sqrt{2m}+
\sqrt{\frac{8\rho_0\pi}{3}}r^{3/2}\right)
\right.\nonumber\\
&&\left.\left[1-\frac{1}{x}{\left(\sqrt{2m}
+\sqrt{\frac{8\rho_0\pi}{3}}r^{3/2}\right)^2}
\right]^{-1}dx \right\} \;,
\end{eqnarray}
where $ J \left( t, x \right) $ is an integrating
factor which always exists and solves the partial differential
equation
\begin{eqnarray}
\partial_{x}J^{-1}
&=&\partial_t\left\{-J^{-1}\frac{1}{\sqrt{x}}
\left(\sqrt{2m}+\sqrt{\frac{8\rho_0\pi}{3}}r^{3/2}\right)
\right.\nonumber\\&&\left.\left[1-\frac{1}{x}{\left(
\sqrt{2m}+\sqrt{\frac{8\rho_0\pi}{3}}r^{3/2}\right)^2}
\right]^{-1}\right\}\;,
\end{eqnarray}
which guarantees that $dT$ is an exact differential. The
metric~(\ref{eq:sch}) is cast in the Schwarzschild form
\begin{eqnarray}
\label{eq:sch1}
ds^2&=&-\left[1-\frac{1}{x}{\left(\sqrt{2m}+\sqrt{
\frac{8\rho_0\pi}{3}}r^{3/2}\right)^2}
\right]J^2dT^2\nonumber\\
&&+\left[1-\frac{1}{x}{\left(\sqrt{2m}
+\sqrt{\frac{8\rho_0\pi}{3}}r^{3/2}\right)^2}
\right]^{-1}dx^2\nonumber\\&&+x^2d\Omega^2
\end{eqnarray}
and the equation of the AHs is
\begin{eqnarray}
\label{eq:eoa}
1-\frac{1}{\sqrt{x}}\left(\sqrt{2m}+
\sqrt{\frac{8\rho_0\pi}{3}}r^{3/2}\right)=0\;.
\end{eqnarray}
Substituting eq.~(\ref{eq:XXXX}) into eq.~(\ref{eq:eoa}),
we obtain
\begin{eqnarray}
\label{eq:seoa}
2\sqrt{6\rho_0\pi}x^{3/2}-
\sqrt{x}\left(3+3\sqrt{6\pi\rho_0}t\right)
+3\sqrt{2m}=0\;.
\end{eqnarray}
This  cubic equation for $\sqrt{x}$ has, in general, only two
positive roots which represent the black hole AH and
the cosmic AH, respectively. In fact, if $\rho_0=0$,
we obtain from eq.~(\ref{eq:seoa})
\begin{eqnarray}
\label{eq:eoasch} x=2m\;,
\end{eqnarray}
the AH of the Schwarzschild black hole. On
the other hand, if $m=0$, it is
\begin{eqnarray}
x=\frac{3}{2} \left( \sqrt{\frac{1}{6\pi\rho_0}}+  t
\right) \;,
\end{eqnarray}
which is the AH of a dust-dominated universe.  The black hole and
the cosmic AHs are plotted in Fig.~\ref{fig:x-t-dust}. As is clear
from this  figure, the radius of the black hole AH
decreases while
that of the cosmic AH increases as the universe evolves. There was
an early time at which the two horizons coincided, and before which
both horizons were absent and the  singularity was naked. Since the
MS mass is proportional to the radius of the black hole AH,  the MS
mass of a black hole decreases with the evolution of the universe.
The reason for this behavior  can be understood as follows: with the
expansion of the universe, the cosmic density is decreasing, which
is equivalent to a decreasing of black hole AH, so the black hole MS
mass decreases.

\begin{figure}
\includegraphics[width=6.5cm]{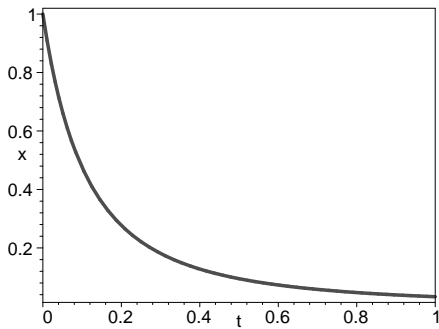}
\caption{The AH of an astronomical black hole in a
{dust-dominated} universe decreases with the cosmic expansion. The
plots correspond to the parameter value $m=10^7
\textrm{M}_{\odot}$. The units of $x$ and $t$ are $2m$ (the
corresponding Schwarzschild radius) and $ \textrm{H}_0^{-1}$,
respectively.} \label{fig:x-t-dust-astro}
\end{figure}

We note that Fig.~\ref{fig:x-t-dust} is plotted {in Plank units
$G=c=\hbar=1$~}. {Translating into International units and taking}
into account the present cosmic density
$\rho_0=10^{-123}~\rho_{P}$ ({where} $\rho_P$ is the Plank
density), we have a very supermassive black hole
$m=10^{24}~\textrm{M}_{\odot}$ for Fig.~\ref{fig:x-t-dust}. The
corresponding units of $x$ and $t$ are $10^{5}~\textrm{Mpc}$ and
the Hubble time $H_0^{-1}$, respectively. To our knowledge, such
supermassive black holes may not exist in the universe, {so} in
Fig.~\ref{fig:x-t-dust-astro} we also plot the evolution of the AH
for an astronomical black hole with mass
$m=10^{7}~\textrm{M}_{\odot}$. We have shifted the Big Bang time
to $t=0$.

{The naked singularity is present before the appearance of the two
apparent horizons and cannot arise from regular  initial data.}

\section{Black hole in a $\Lambda\textrm{CDM}$ universe}

Astronomical observations show that the present-day universe is
dominated by dark energy \cite{Per:99,Rie:98}, hence we
seek  exact solutions  describing  a black hole immersed in a
mixture of dark matter
 and dark energy. The corresponding stress-energy tensor is
\begin{eqnarray}
\label{eq:emt} T_{\mu\nu}=\rho_d
U_{\mu}U_{\nu}+\rho_{\Lambda}g_{\mu\nu}\;.
\end{eqnarray}
The first term on  the right hand side describes dark matter and
the second term describes dark energy. $\rho_{\Lambda}$ is a
positive constant \footnote{ {There is evidence in \cite{Kal:2010}
and  \cite{lake:2011} that the presence of a cosmological
constant makes a significant difference in the McVittie solution,
and we expect that this is the case also for our  generalized
Lema\^itre-Tolman-Bondi solutions.} } We assume the  line element
to be of the form
\begin{eqnarray}
\label{eq:line} ds^2=-dt^2+e^{\bar
\omega}dr^2+e^{\omega}d\Omega^2\;,
\end{eqnarray}
where $\bar{\omega}$ and $ \omega$ are functions of $t$ and $ r$. In
comoving coordinates, the four-velocity is $U^{\mu}=(1,
0, 0, 0)$ and the Einstein equations are
\begin{eqnarray}
\label{eq:phantomEin}
G_0^0&=&8\pi\left(\rho_d+\rho_{\Lambda}\right)\;, \ \ \
G_1^0 =0\;,\nonumber\\G_1^1&=&8\pi \rho_{\Lambda}\;, \ \ \ G_2^2=8\pi
\rho_{\Lambda}\;.
\end{eqnarray}
From $G_1^0=0$ one obtains
\begin{eqnarray}
\label{eq:mphantomw}e^{\bar{\omega}}=e^{\omega}{\omega'^{2}}/4\;.
\end{eqnarray}
Note that there is no accretion onto the central object, as in the
McVittie metric. Following Oppenheimer and Snyder, we have set the
integration  ``constant'' ($f(R)$ in \cite{op:1939}) to  unity.
Substituting eq.~(\ref{eq:mphantomw}) into
eq.~(\ref{eq:phantomEin}),
we obtain
\begin{eqnarray}
\label{eq:pww}
\ddot{\omega}'+\frac{3}{2} \,
\dot{\omega}\dot{\omega}'=0\;.
\end{eqnarray}
Integration with respect to $r$ yields
\begin{eqnarray}
 \ddot{\omega}+\frac{3}{4}\dot{\omega}^2=-8\pi
K\left(t\right)\;,
\end{eqnarray}
where $K$ is an integration ``constant'' (an arbitrary function of $t$).
This equation is consistent with  $G_1^1=8\pi
\rho_{\Lambda}$ if and only if
\begin{eqnarray}
\label{eq:ppre}
K\left(t\right)=-\rho_{\Lambda}\;.
\end{eqnarray}
The Einstein equations then reduce to
\begin{eqnarray}
\label{eq:pein}
G_1^1= \ddot{\omega}+\frac{3}{4}\dot{\omega}^2&=&8\pi
\rho_{\Lambda}\;,\nonumber \\
G_0^0=
\frac{\dot{\omega}'\dot{\omega}}{\omega'}+\frac{3}{4}\dot{\omega}^2&=&8\pi
\left(\rho_d+\rho_{\Lambda}\right)\;.
\end{eqnarray}

Using $k\equiv \sqrt{6\pi\rho_{\Lambda}}$, the Einstein
equations have the solution
\begin{eqnarray}
\label{eq:ww}
e^{\omega}&=&\left(Se^{kt}+Pe^{-kt}\right)^{4/3}\;,\nonumber\\
\nonumber\\
e^{\bar{\omega}}&=&e^{\omega}\omega'^2/4\;,
\end{eqnarray}
where $S$ and $ P$ are arbitrary functions of $r$. Substitution
of eqs.~({\ref{eq:ww}}) into eq.~(\ref{eq:line}) and the
use of $S$
as a new spatial coordinate leave only one degree of freedom. It is
convenient to choose
\begin{eqnarray}
\label{eq:S} S=r^{3/2}\;,
\end{eqnarray}
which has the advantage  that if $P=0$ and $ \rho_{\Lambda}=0$, the
line element~(\ref{eq:line}) reduces to the Minkowski one.

\subsection{The Schwarzschild-de Sitter solution}

The Einstein equations with $\rho_d=0$ yield
\begin{eqnarray}
\label{eq:sch-de}P=S_0r^{-3/2}\;,
\end{eqnarray}
where $S_0$ is an integration constant. The Schwarzschild-de Sitter
solution is given by
\begin{eqnarray}
\label{eq:schde} e^{\omega}&=&\left(r^{3/2}e^{kt}+S_0r^{-3/2}
e^{-kt}\right)^{4/3}\;,\nonumber\\
\nonumber\\
e^{\bar{\omega}}&=&e^{\omega}\omega'^2/4\;.
\end{eqnarray}
The physical meaning of $S_0$ can be understood after rewriting
eq.~(\ref{eq:schde}) in Schwarzschild coordinates; setting
\begin{eqnarray}
\label{eq:xx} x&=&\left(r^{3/2} e^{kt}+{S_0}r^{-3/2}
e^{-kt}\right)^{2/3}\;,
\end{eqnarray}
the line element~(\ref{eq:line}) becomes
\begin{eqnarray}
\label{eq:schde1}
ds^2&=&-\left[1-\frac{4k^2\left(x^3-4S_0
\right)}{9x}\right]dt^2+dx^2\nonumber\\
&&\nonumber\\
&&
-\frac{4k}{3\sqrt{x}}\sqrt{x^3-4S_0} \,
dtdx+x^2d\Omega^2\;.
\end{eqnarray}
Introducing the new time
\begin{eqnarray}
\label{eq:newtime}
T&=&t+\int\left[1-\frac{4k^2\left(x^3-4S_0\right)}{9x}
\right]^{-1}\nonumber\\&&\frac{2k}{3\sqrt{x}}
\sqrt{x^3-4S_0} \, dx\;,
\end{eqnarray}
the metric~(\ref{eq:schde1}) simplifies to
\begin{eqnarray}
\label{eq:schde2}
ds^2&=&-\left(1+\frac{16k^2S_0}{9x}-\frac{4k^2}{9}x^2
\right)dT^2+x^2d\Omega^2\nonumber\\
&&\nonumber\\
&&
+\left(1+\frac{16k^2S_0}{9x}-\frac{4k^2}{9}x^2\right)^{-1}dx^2\;.
\end{eqnarray}
Comparing eq.~(\ref{eq:schde2}) with the well-known form of
the Schwarzschild-de Sitter solution
\begin{eqnarray}
\label{eq:schde3}
ds^2&=&-\left(1-\frac{2m}{x}-
\frac{8\pi\rho_{\Lambda}}{3}x^2\right)dT^2+x^2d\Omega^2\nonumber\\&&
+\left(1-\frac{2m}{x}-\frac{8\pi\rho_{\Lambda}}{3}x^2\right)^{-1}dx^2\;;
\end{eqnarray}
one deduces the physical meaning of $S_0$,
\begin{eqnarray}
\label{eq:s0} S_0&=&-\frac{9m}{8k^2}\;.
\end{eqnarray}
Then, the Schwarzschild-de Sitter solution in comoving coordinates is
given by
\begin{eqnarray}
\label{eq:schde4}
e^{\omega}&=&\left(r^{3/2}e^{kt}-\frac{9m}{8k^2}r^{-3/2}
e^{-kt}\right)^{4/3}\;,\nonumber\\
\nonumber\\
e^{\bar{\omega}}&=&e^{\omega}\omega'^2/4\;.
\end{eqnarray}

\subsection{$\Lambda \textrm{CDM}$ universe}

Eqs.~(\ref{eq:ww}) and~(\ref{eq:S}) show that we recover the
homogenous and isotropic, spatially flat universe with a mixture of
dust matter and dark energy if and only if we
set $P=-\frac{3\rho_0}{16k^2}r^{3/2}$. This solution is obtained for
\begin{eqnarray}
\label{eq:universe}
e^{\omega}&=&\left(r^{3/2}e^{kt}-\frac{3\rho_0}{16k^2}r^{3/2}
e^{-kt}\right)^{4/3}\;,\nonumber\\
\nonumber\\
e^{\bar{\omega}}&=&e^{\omega}\omega'^2/4\;.
\end{eqnarray}
The density of dust/dark matter  and the scale factor are
given by
\begin{eqnarray}
\label{eq:dustDE} \rho_d=\frac{\rho_0}{a^3},\ \ \ \
a=\left(e^{kt}-\frac{3\rho_0}{16k^2}e^{-kt}\right)^{2/3}\;,
\end{eqnarray}
where $t>\frac{1}{2k}\ln\frac{3\rho_0}{16k^2}$, and where
$t=\frac{1}{2k}\ln\frac{3\rho_0}{16k^2}$ represents the
Big Bang
singularity.

\subsection{Black hole in a $\Lambda \textrm{CDM}$ universe}

The inspection of eqs.~(\ref{eq:schde4})
and~(\ref{eq:universe}) suggests
that  the solution for a black hole in a $\Lambda \textrm{CDM}$
universe should be given by
\begin{eqnarray}
\label{eq:bhlambda}
e^{\omega}&=&\left(r^{3/2}e^{kt}-\frac{9m}{8k^2}r^{-3/2}
e^{-kt}-\frac{3\rho_0}{16k^2}r^{3/2}
e^{-kt}\right)^{4/3}\;,\nonumber\\
\nonumber\\
e^{\bar{\omega}}&=&e^{\omega}\omega'^2/4\;.
\end{eqnarray}
It is clear, by comparison of eqs.~(\ref{eq:bhlambda})
and~(\ref{eq:ww}), that this is indeed a solution.  The substitution
of eq.~(\ref{eq:bhlambda}) into the Einstein equations
gives the
density of dust
\begin{eqnarray}
\label{eq:dustden} \rho_d=\frac{\rho_0\left(32r^3e^{kt}+6m
e^{-kt}-\rho_0r^3e^{-kt}\right)^2}{R^3\left(32r^3e^{kt}-6m
e^{-kt}-\rho_0r^3e^{-kt}\right)^2} \;.
\end{eqnarray}
This quantity is positive on the entire spacetime manifold. If $m=0$,
we have $\rho_d=\rho_0/R^3$, the usual scaling.

\section{Black Hole Evolution in a $\Lambda \textrm{CDM}$
universe}

Here we investigate the evolution of a black hole AH in a
$\Lambda \textrm{CDM}$ universe. For this purpose, we  rewrite the
metric in Schwarzschild coordinates using
\begin{eqnarray}
\label{eq:xxx} x& \equiv &e^{-2kt/3}r\left(e^{2kt}-\frac{9m}{8k^2r^3}
-\frac{3\rho_0}{16k^2}\right)^{2/3}\;,
\end{eqnarray}
eq.~(\ref{eq:line}) becomes
\begin{eqnarray}
\label{eq:schdedu}ds^2&=&
-\left(1-\frac{2m}{x}-\frac{4k^2}{9}x^2-
\frac{\rho_0}{3x}r^3\right)dt^2\nonumber\\
&&\nonumber \\
&&
-2\left(\frac{2m}{x}+\frac{4k^2}{9}x^2+\frac{\rho_0}{3x}
r^3\right)^{1/2}dtdx\nonumber\\
&&\nonumber \\
&&+dx^2+x^2d\Omega^2\;.
\end{eqnarray}
Introducing the new time variable $T$ with
\begin{eqnarray}
\label{eq:newT}
dT&=&{J}^{-1}dt+{J}^{-1}\left(1-\frac{2m}{x}
-\frac{4k^2}{9}x^2-\frac{\rho_0}{3x}r^3\right)^{-1}
\nonumber\\
&&\nonumber\\
&&
\cdot\left(\frac{2m}{x}+\frac{4k^2}{9}x^2+
\frac{\rho_0}{3x}r^3\right)^{1/2}dx\;,
\end{eqnarray}
where $J(t,x)$ is  an integrating factor, we obtain a
black hole solution in a $\Lambda \textrm{CDM}$ universe.
\begin{eqnarray}
\label{eq:schdedu1}ds^2&=&-\left(1-\frac{2m}{x}-
\frac{4k^2}{9}x^2-\frac{\rho_0}{3x}r^3\right)J^2dT^2
+x^2d\Omega^2\nonumber\\
&&\nonumber \\
&&
+\left(1-\frac{2m}{x}-\frac{4k^2}{9}x^2
-\frac{\rho_0}{3x}r^3\right)^{-1}dx^2\;.
\end{eqnarray}
The equation  of the AHs is
\begin{eqnarray}
\label{eq:EOA}
g_{00}=1-\frac{2m}{x}-\frac{4k^2}{9}x^2-\frac{\rho_0}{3x}r^3=0 \;,
\end{eqnarray}
so
 \begin{eqnarray}
 \label{eq:rrr}
r&=&\left(\frac{3x}{\rho_0}-
\frac{6m}{\rho_0}-\frac{4k^2x^3}{3\rho_0}\right)^{1/3}.
\end{eqnarray}
The substitution of eq.~(\ref{eq:rrr}) into
eq.~(\ref{eq:xxx}) yields
the equation of the AHs,  using which the AHs evolution is
plotted  in Fig.~3. The black hole AH shrinks while the
cosmic AH expands with the expansion of the
universe. There was a time in the past at which the two horizons coincided
and before which no horizon existed  and the
singularity was naked. Late in the history of the
universe, both black hole  and cosmic AHs approach constant values.

\begin{figure}
\includegraphics[width=6.5cm]{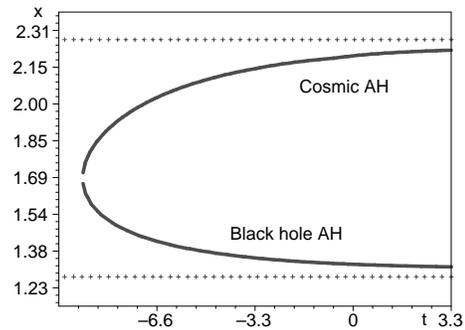}
\caption{In a $\Lambda$CDM universe, the black hole AH
(lower {\bf curve}) shrinks while the cosmic AH (upper {\bf
curve}) expands with the cosmic expansion. There was a time
at which the two horizons coincided and before which no
horizon existed  and the singularity was naked. Late in the
history of the universe, both black hole and cosmic AHs
approach constant size. The plots correspond to the
parameter values $m=2.2\cdot10^{22}M_{\bigodot},
\rho_0=0.27\rho_{C}, \rho_{\Lambda}=0.73\rho_C$. $\rho_C$
is the current cosmic density. The unit of $t$ is the
present Hubble time $H_0^{-1}$. }
\label{fig:x-t}
\end{figure}

\section{Black hole in a quintom universe}

The first year WMAP data combined with the \textrm{2dF} galaxy
survey and the supernova Ia data favor the phantom energy equation
of state of the cosmic fluid $w<-1$ over the cosmological
constant ($w=-1$)  and the quintessence field ($w>-1$)
\cite{ps:04,alam:04}. The
data seem to slightly favor an evolving dark energy with $w<-1$ at
the present epoch and $w>-1$ in the near past \cite{fe:04}. The dark
energy candidate with evolving $w$ is named \emph{quintom}
\cite{feng:2004}. In this section, we look for the solution for a
black hole in a quintom-dominated universe.  We take the energy-momentum
tensor for the quintom fluid to be of the form
\begin{eqnarray}
\label{eq:phantom} T_{\mu\nu}=\left(\rho+p\right)
U_{\mu}U_{\nu}+pg_{\mu\nu}\;,
\end{eqnarray}
and the line element in the form of
eq.~(\ref{eq:line}). In comoving coordinates, the
four-velocity is $U^{\mu}=(1, 0, 0, 0)$ and the
Einstein equations are
\begin{eqnarray}
\label{eq:quintomEin} G_0^0&=&8\pi\rho\;, \ \ \
G_1^0 =0\;,\nonumber\\G_1^1&=&-8\pi p\;, \ \ \ G_2^2=-8\pi p\;.
\end{eqnarray}
From $G_1^0=0$ (again, there is no accretion), we obtain
\begin{eqnarray}
\label{eq:quintomw}e^{\bar{\omega}}=e^{\omega}{\omega'}^2/4\;.
\end{eqnarray}
Again, we set the integration
``constant'' to  unity and, inserting
eq.~(\ref{eq:quintomw}) into eqs.~(\ref{eq:quintomEin}), we
obtain
\begin{eqnarray}
\ddot{\omega}'+\frac{3}{2} \,
\dot{\omega}\dot{\omega}'=0\;,
\end{eqnarray}
and
\begin{eqnarray}
 \ddot{\omega}+\frac{3}{4}\dot{\omega}^2=-8\pi
K\left(t\right)\;,
\end{eqnarray}
where $K(t)$ is an integration ``constant''.
This equation is consistent with  $G_1^1=-8\pi p$
only if the pressure satisfies
\begin{eqnarray}
p=K\left(t\right)\;,
\end{eqnarray}
{\em i.e.}, it is spatially  homogenous, which can be seen as the
consequence of  three facts. First,  we have
set $f(R)=1$ for a spatially flat  background. Second,
we use comoving coordinates and, third  and most
important, we have taken the source to be a \emph{single perfect
fluid}. Therefore, the comoving observer will see a homogenous
pressure. The Einstein equations simplify to
\begin{eqnarray}
\label{eq:qein}
 \ddot{\omega}+\frac{3}{4}\dot{\omega}^2&=&-8\pi
p\left(t\right)\;,\nonumber \\
&&\nonumber \\
\frac{\dot{\omega}'\dot{\omega}}{\omega'}+\frac{3}{4}\dot{\omega}^2&=&8\pi
\rho\left(t, r\right)\;.
\end{eqnarray}
Now we have three functions $\omega, p$, and  $\rho$ but only two
equations and the system is  not closed. For simplicity, we
assign the pressure in the form
\begin{eqnarray}
\label{eq:qpp} p=-\frac{p_0}{\left(t_0-t\right)^2}\;,
\end{eqnarray}
where $p_0$ is  a positive constant to keep the
pressure always negative, and $t_0$ is a positive constant
identifying the Big Rip singularity. The general solution of
eqs.~(\ref{eq:qein}) is then given by
\begin{eqnarray}
\label{eq:pgen}
e^{\omega}&=&\left[P\left(r\right)\left(t_0-t\right)^{\frac{1}{2}
\left(1-\sqrt{1+24\pi
p_0}\right)}
\right.\nonumber\\&&\left.+S\left(r\right)
\left(t_0-t\right)^{\frac{1}{2}\left(1+\sqrt{1+24\pi
p_0}\right)}\right]^{4/3}\;,
\end{eqnarray}
where $S$ and $P$ are arbitrary functions of $r$. It is convenient
to choose
\begin{eqnarray}
\label{eq:pS} P=r^{3/2}\;,
\end{eqnarray}
then the {candidate} solution for a black hole in a quintom
universe is
\begin{eqnarray}
\label{eq:bhqu}
e^{\omega}&=&\left[
r^{3/2}\left(t_0-t\right)^{\frac{1}{2}\left(1-\sqrt{1+24\pi
p_0}\right)}\right.\nonumber\\
&&\nonumber \\
&&\left.-\frac{{3}}{2}\sqrt{2m}
\left(t_0-t\right)^{\frac{1}{2}\left(1+\sqrt{1+24\pi
p_0}\right)}\right.\nonumber\\
&&\nonumber \\
&&\left.-\frac{{3}}{2}
\sqrt{\frac{8\pi\rho_0}{3}}r^{3/2}\left(t_0-t\right)^{\frac{1}{2}
\left(1+\sqrt{1+24\pi p_0}\right)}\right]^{4/3}\;.
\end{eqnarray}
If $m=0$, we get a quintom-dominated cosmology while,  if $p_0=0$,
we get {a candidate}  solution for a black hole in a
matter-dominated universe. Substituting eq.~(\ref{eq:bhqu}) into
eqs.~(\ref{eq:qein}), we find that the energy density is positive
and inhomogeneous.

\section{Evolution of a black hole in a quintom universe}

\begin{figure}
\includegraphics[width=6.5cm]{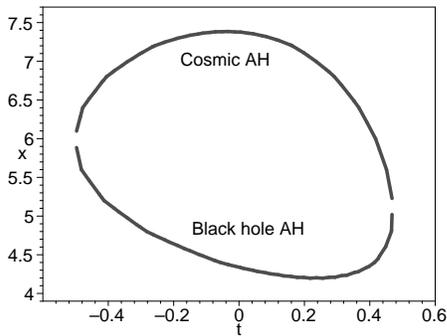}
\caption{In a quintom-dominated universe, the black hole AH (lower
{curve}) first shrinks and then expands, while the cosmic AH
(upper {curve}) first expands and then shrinks. There exist two
instants of time, one in the past and one in the future, at which
the two horizons coincide. Before or after these critical
instants, no AH exists and a naked singularity is present. The
plots correspond to the parameter values $m=1,\rho_0=0.0002,
p_0=0.001,$ and $  t_0=1$.} \label{fig:x-t-quintom}
\end{figure}

To investigate the evolution of the black hole AH, we rewrite the
metric in Schwarzschild coordinates using
\begin{eqnarray}
\label{eq:qbx}x& \equiv &\left[
r^{3/2}\left(t_0-t\right)^{\frac{1}{2}\left(1-\sqrt{1+24\pi
p_0}\right)}\right.\nonumber\\
&&\nonumber \\
&&\left.
-\frac{{3}}{2}\sqrt{2m}\left(t_0-t\right)^{\frac{1}{2}\left( 1+\sqrt{1+24\pi
p_0}\right)}\right.\nonumber\\
&&\nonumber \\
&&\left.-\frac{{3}}{2}
\sqrt{\frac{8\pi\rho_0}{3}}r^{3/2}\left(t_0-t\right)^{
\frac{1}{2}\left(1+\sqrt{1+24\pi
p_0}\right)}\right]^{2/3}\;,
\end{eqnarray}
and define $k\equiv \sqrt{1+24\pi p_0}$.
Eq.~(\ref{eq:line}) becomes
\begin{eqnarray}
\label{eq:line3}ds^2&=&-\left(1-\mathscr{L}^2\right)dt^2
+dx^2-2\mathscr{L}dtdx+x^2d\Omega^2\;,\nonumber\\
\mathscr{L}&\equiv& \frac{\partial x}{\partial t}.
\end{eqnarray}
Introducing the time coordinate
\begin{eqnarray}
dT&=&J^{-1}\left(dt+
\frac{\mathscr{L}}{1-\mathscr{L}^2}dx\right)\;,
\end{eqnarray}
where $J$ is an integrating factor, the metric
becomes
\begin{eqnarray}
\label{eq:line4}ds^2&=&-\left(1-\mathscr{L}^2\right)
J^2dt^2+\frac{1}{1-\mathscr{L}^2}dx^2+x^2d\Omega^2
\end{eqnarray}
and the AHs are identified by
\begin{eqnarray}
\label{eq:qhAH}1-\mathscr{L}^2=0\;.
\end{eqnarray}
Eq.~(\ref{eq:qbx}) then yields
\begin{eqnarray}
\label{eq:qrrrr}r^{3/2}&=&\frac{1}{2}\frac{2x^{3/2}+
3\sqrt{2m}\left(t_0-t\right)^{\frac{1}{2}\left(1+k\right)}}
{\left(t_0-t\right)^{\frac{1}{2}\left(1-k\right)}
-\sqrt{6\pi\rho_0}\left(t_0-t\right)^{\frac{1}{2}\left(1+k\right)}}\;.
\end{eqnarray}
Substituting eq.~(\ref{eq:qrrrr}) into eq.~(\ref{eq:qhAH}), an
explicit equation for the AHs in terms of $t$ and $x$ is obtained.
The evolution of the AHs is reported in
Fig.~\ref{fig:x-t-quintom}.  In the quintom-dominated universe,
the size of the black hole AH first decreases in the matter era
and then  increases during the phantom-dominated epoch
\cite{caldwell:2003}. The size of the cosmic AH, instead, first
increases and then  decreases. There exists an instant of time in
the past at which the two horizons coincided. Both horizons
disappear before the Big Rip, leaving behind a  naked singularity
and violating Cosmic Censorship  \cite{pen:1969}. This result is
consistent with our previous discussion in \cite{gao:2008}. {The
naked singularity does not arise from regular Cauchy data, in fact
it was already present before the appearance of the two apparent
horizons.}

The reason for the increase of the black hole mass  in the
phantom-dominated epoch can be understood as follows: the density of
the universe is increasing, which makes the black hole AH also
increase. Hence, the black hole mass (one half of the AH radius)
increases. On
the contrary, in a matter-dominated universe, the cosmic density is
decreasing, so the black hole AH and the black hole mass decrease. In a
cosmological constant-dominated universe, the cosmic density is
constant and the black hole mass also stays constant.

\section{Discussion and conclusions}

We have shown that several well-known and less-known solutions of
the Einstein equations can not describe a black hole embedded in a
matter-dominated universe sourced by a single perfect fluid. This
includes the McVittie, Thakurta, Sultana-Dyer, Vaidya, and
Faraoni-Jacques solutions. Motivated by this fact, we  have
constructed exact solutions of the Einstein equations {purporting
to represent} a black hole embedded in a background universe which
are as simple as possible, beginning with solutions with a single
perfect fluid. These new solutions generalize the
Lema\^{\i}tre-Tolman-Bondi metrics (which are restricted to dust
universes) and do not exhibit accretion of the surrounding cosmic
fluid onto the central black hole. We have found metrics {which
presumably describe} a black hole immersed in a matter-dominated,
a matter plus dark energy-dominated universe, and a
quintom-dominated universe, respectively.

The AH and the Misner-Sharp mass are related to the
thermodynamics of dynamical black holes and FRW universes
\cite{ward:2000,bak:99,cai:05,recent:2008,
gong:07,ak:07,caic:07,caicao:07,akc:07,she:07,shey:07,DiCriscienzoetal07},
and, therefore, we investigate the evolution of the black hole and
cosmic AHs, as well as  the Misner-Sharp mass. We find that the
black hole mass decreases in a matter- or  matter plus
cosmological constant-dominated universe. In a quintom-dominated
universe, the black hole mass decreases in the matter era, while
it increases in the phantom-dominated epoch.  The physical reason
is that the cosmic density  first decreases and then increases,
which makes the radius of the black hole AH first decreasing and
then increasing. Then, also the black hole mass (one half of the
AH radius) first decreases and then increases. An interesting
result is that the AHs will disappear and the singularity will
become naked before the  Big Rip is reached, in violation of the
Cosmic Censorship  Conjecture \cite{pen:1969}. If the latter is
correct, then phantom matter may not be allowed to exist in
nature. This statement, however, must be taken with a grain of
salt: in fact, it is based on a particular solution of the field
equations which may be very special. While the result is still
interesting because very few exact solutions are known to
represent the physical situation under study, the solution
proposed may still be too special, or fine-tuned, to draw general
conclusions. The phenomenology reported here, however, matches
that obtained with very different classes of solutions (which are
indeed accreting cosmic fluid) in \cite{gao:2008}.

Another rather surprising result is that the black hole mass
decreases in a matter-dominated universe. As the most strongly
bound gravitational system, it seems intuitive that  a black hole
will have its mass increasing because of the swallowing of
surrounding cosmic matter. Our solution represents cases in which
there is no accretion onto the central black hole (as described by
$G_1^0=0$) and in which the  cosmic  expansion ``wins'' over the
local gravitational attraction. The universe is always expanding
and the density of the cosmic fluid always decreases with the
cosmic time. The seemingly bizarre behavior of the MS mass derives
from the fact that it is really a mass sum ({\em  i.e.}, the mass
of the background fluid is included), {and it coincides with the
Hawking-Hayward quasi-local mass} \cite{gao:2008}. In the absence
of accretion of cosmic fluid onto the black hole, this mass sum
decreases in any universe in which the cosmological fluid
satisfies the weak energy condition and its energy density
decreases with the cosmic expansion. In the presence of phantom
matter (such as in the late epoch of a quintom-dominated
universe), the weak energy condition is violated, the phantom
energy becomes more concentrated, instead of being diluted,  with
the cosmic expansion, and the mass sum increases.

Future work will concentrate on studying the generality of the
solutions presented here, and of previous solutions {interpreted
as} black holes embedded in a cosmological space. {To fully
support  the interpretation of these solutions as black holes in a
cosmological background}, it is {important that} a detailed
investigation  on the global structure of these spacetimes  be
{provided. Until this global analysis is available, we cannot yet
claim  that our solutions do represent black holes embedded in
expanding  universes, although there is circumstantial evidence.
This  analysis will be carried out in future publications.}

\section{Acknowledgments}

We thank J. D. Barrow, A. Krasinski and J. Taghizade for helpful
communications. We also thank {an} anonymous referee for expert
and insightful comments, which {led to improvements in this
manuscript}. This work is supported by the National Science
Foundation of China under the Key Project Grant 10533010, Grant
10575004, Grant~10973014, and the 973 Project (No. 2010CB833004).
V.F. acknowledges the Natural Sciences and Engineering Research
Council of Canada (NSERC).

After completion of this work, we learned that the solutions of
Secs. V and VI can also be obtained as special cases of solutions
studied by Krasinski, Hellaby, and Jacewicz (see Ref. [3]). The
solution of Secs. IX and X was already contained in Ref. [64] and
is analogous to class-II Szekeres models. We thank A. Krasinski
for pointing this out.

\newcommand\AL[3]{~Astron. Lett.{\bf ~#1}, #2~ (#3)}
\newcommand\AP[3]{~Astropart. Phys.{\bf ~#1}, #2~ (#3)}
\newcommand\AJ[3]{~Astron. J.{\bf ~#1}, #2~(#3)}
\newcommand\APJ[3]{~Astrophys. J.{\bf ~#1}, #2~ (#3)}
\newcommand\APJL[3]{~Astrophys. J. Lett. {\bf ~#1}, L#2~(#3)}
\newcommand\APJS[3]{~Astrophys. J. Suppl. Ser.{\bf ~#1}, #2~(#3)}
\newcommand\JCAP[3]{~JCAP. {\bf ~#1}, #2~ (#3)}
\newcommand\LRR[3]{~Living Rev. Relativity. {\bf ~#1}, #2~ (#3)}
\newcommand\MNRAS[3]{~Mon. Not. R. Astron. Soc.{\bf ~#1}, #2~(#3)}
\newcommand\MNRASL[3]{~Mon. Not. R. Astron. Soc.{\bf ~#1}, L#2~(#3)}
\newcommand\NPB[3]{~Nucl. Phys. B{\bf ~#1}, #2~(#3)}
\newcommand\PLB[3]{~Phys. Lett. B{\bf ~#1}, #2~(#3)}
\newcommand\PRL[3]{~Phys. Rev. Lett.{\bf ~#1}, #2~(#3)}
\newcommand\PR[3]{~Phys. Rep.{\bf ~#1}, #2~(#3)}
\newcommand\PRD[3]{~Phys. Rev. D{\bf ~#1}, #2~(#3)}
\newcommand\RMP[3]{~Rev. Mod. Phys.{\bf ~#1}, #2~(#3)}
\newcommand\SJNP[3]{~Sov. J. Nucl. Phys.{\bf ~#1}, #2~(#3)}
\newcommand\ZPC[3]{~Z. Phys. C{\bf ~#1}, #2~(#3)}

\end{document}